\documentclass[12pt,reqno]{amsart}
\usepackage{amsmath,amsthm,amssymb,comment,fullpage}
\usepackage{braket}
\usepackage{mathtools}
\usepackage{caption}
\usepackage{times}
\usepackage[T1]{fontenc}
\usepackage{mathrsfs}
\usepackage{latexsym}
\usepackage[dvips]{graphics}
\usepackage{epsfig}
\usepackage{amsmath,amsfonts,amsthm,amssymb,amscd}
\usepackage{tikz}
\usepackage{tkz-graph}
\usepackage{enumerate}
\tikzstyle{vertex}=[circle, draw, inner sep=0pt, minimum size=6pt]

\tikzstyle{vtx}=[circle, draw, inner sep=0pt, minimum size=12pt]

\usepackage{color}
\usepackage{hyperref}
\usepackage{url}

\usepackage[norelsize]{algorithm2e}

\title{Phylogenetic Trees}

\author{Hector Ba\~nos}
\email{\textcolor{blue}{\href{mailto:hdbanoscervantes@alaska.edu}{hdbanoscervantes@alaska.edu}}}
\address{Department of Mathematics and Statistics, University of Alaska Fairbanks, Fairbanks, AK 99775}

\author{Nathaniel Bushek}
\email{\textcolor{blue}{\href{mailto:nbushek@alaska.edu}{nbushek@alaska.edu}}}
\address{Department of Mathematics and Statistics, University of Alaska Anchorage, Anchorage, AK 99508}

\author{Ruth Davidson}
\email{\textcolor{blue}{\href{mailto:redavid2@illinois.edu}{redavid2@illinois.edu}}}
\address{Department of Mathematics, University of Illinois Urbana-Champaign, Urbana, IL 61801}

\author{Elizabeth Gross}
\email{\textcolor{blue}{\href{mailto:elizabeth.gross@sjsu.edu}{elizabeth.gross@sjsu.edu}}}
\address{Department of Mathematics, San Jose State University, San Jose, CA 95192}

\author{Pamela E. Harris}
\email{\textcolor{blue}{\href{mailto:pamela.e.harris@williams.edu}{pamela.e.harris@williams.edu}}}
\address{Department of Mathematics and Statistics, Williams College, Williamstown, MA 01267}

\author{Robert Krone}
\email{\textcolor{blue}{\href{mailto:rk71@queensu.ca}{rk71@queensu.ca}}}
\address{Department of Mathematics and Statistics, Queens University, Kingston, ON K7L 3N6, Canada}

\author{Colby Long}
\email{\textcolor{blue}{\href{mailto:celong@ncsu.edu}{celong@ncsu.edu}}}
\address{Department of Mathematics, North Carolina State University, Raleigh, NC 27695}

\author{Allen Stewart}
\email{\textcolor{blue}{\href{mailto:stewaral@seattleu.edu}{stewaral@seattleu.edu}}}
\address{Department of Mathematics, Seattle University, Seattle, WA 98122}

\author{Robert Walker}
\email{\textcolor{blue}{\href{mailto:robmarsw@umich.edu}{robmarsw@umich.edu}}}
\address{Department of Mathematics, University of Michigan, Ann Arbor, MI 48109}

\subjclass[2010]{Primary 13P25, Secondary 14M25, 05C05, 92D15}

\keywords{Phylogenetic Trees, Toric Ideals, Secant Ideals}

\date{\today}

\begin{document}

\maketitle

\begin{abstract}
We introduce the package {\it PhylogeneticTrees} for {\it Macaulay2} which allows users to compute phylogenetic invariants for group-based tree models. We provide some background information on phylogenetic algebraic geometry and show how the package {\it PhylogeneticTrees} can be used to calculate a generating set for a phylogenetic ideal as well as a lower bound for its dimension. Finally, we show how methods within the package can be used to compute a generating set for the join of any two ideals.
\end{abstract}

$\ $\\
$\ $\\
\noindent
{\bf Motivation.} A central problem in phylogenetics is to describe the evolutionary history of $n$ species from their aligned DNA sequences. One way to do this is through a model-based approach. 
A phylogenetic model is a statistical model, specified parametrically, of molecular evolution at a single DNA site.  
We can regard the aligned sequences as a collection of $n$-tuples of the four DNA bases, one from each site. Each choice of parameters results in a probability distribution on the $4^n$ possible $n$-tuples.
The goal of model-based reconstruction is to find a choice of parameters that yields a distribution close to the empirical distribution. If we are able to do so then it is reasonable to assume that the model is an accurate reflection of the 
underlying evolutionary process. Most significantly, we can infer that the underlying tree parameter of the phylogenetic model is the evolutionary tree of the species under consideration.\\

\noindent
{\bf Mathematical Background.} In phylogenetic algebraic geometry, the statistical models under consideration are tree-based Markov models.
 This means that we assume a Markov process proceeds along a  tree with a transition matrix associated to each edge. A $\kappa$-state phylogenetic model on an $n$-leaf tree $\mathcal{T}$ induces a polynomial map
from the parameter space $\Theta_{\mathcal{T}} \subseteq \mathbb{R}^m$ to the probability simplex $\Delta^{\kappa^n -1}$
\[
\psi_{\mathcal{T}}: \Theta_{\mathcal{T}} \rightarrow \Delta^{\kappa^n -1} \subset \mathbb{C}^{\kappa^n}.
\]
The image of this map is the set of all probability distributions we obtain by varying the entries of the transition matrices; we refer to the image as the model $\mathcal{M}_\mathcal{T}$. For phylogenetic applications, usually $\kappa=2$ or $\kappa=4$.

The Zariski closure of the model
$V_\mathcal{T} :=
\overline{\mathcal{M}_\mathcal{T}} \subseteq 
\mathbb{C}^{\kappa ^ n}$
is an affine algebraic variety.
For the models we consider, the entries of $\psi_{\mathcal{T}}$ are homogeneous polynomials of uniform degree. Thus, $V_{\mathcal{T}}$ can be viewed as the affine cone of a
projective variety in 
$\mathbb{P}^{\kappa^n-1}$.
The ideal $\mathcal{I}_{\mathcal{T}} := 
I(V_{\mathcal{T}}) \subseteq \mathbb C[x_1, \ldots, x_{\kappa^n}]$ of all polynomials vanishing on $V_{\mathcal T}$ 
is a homogeneous ideal called the \textit{ideal of 
phylogenetic invariants}, this ideal carries useful information about the model that can be used for determining model identifiability and performing model selection 
\cite{APRS2011, Allman2012, CavenderFelsenstein1987, CFS2006, Lake1987,  Long2015a, Matsen2008, Matsen2007, Rhodes2012, Rus2012}.

Given two models, $\mathcal{M}_{\mathcal{T}_1}$
and $\mathcal{M}_{\mathcal{T}_2}$, 
we define the \textit{mixture model} 
$\mathcal{M}_{\mathcal{T}_1}*\mathcal{M}_{\mathcal{T}_2}$ to be the 
image of the map
\begin{equation}\label{mixturedef}
    \psi_{\mathcal{T}_1, \mathcal{T}_2}: \Theta_{\mathcal{T}_1} \times \Theta_{\mathcal{T}_2} \times [0,1] \rightarrow \Delta^{\kappa^n -1} \subset \mathbb{C}^{\kappa^n};
  \;\;\;\;\; \psi_{\mathcal{T}_1, \mathcal{T}_2}(\theta_1, \theta_2, \pi) = \pi  \psi_{\mathcal{T}_1}(\theta_1) + (1- \pi)  \psi_{\mathcal{T}_2}(\theta_2). 
\end{equation}
As before, we take the Zariski closure $\overline{
\mathcal{M}_{\mathcal{T}_1}*
\mathcal{M}_{\mathcal{T}_2}} \subseteq 
 \mathbb{C}^{\kappa^n}$, and now we obtain the algebraic variety 
$V_{\mathcal{T}_1 * \mathcal{T}_2} =
V_{\mathcal{T}_1} * V_{\mathcal{T}_2}$,
the join of
$V_{\mathcal{T}_1}$ and 
$V_{\mathcal{T}_2}$.
The \emph{join} of two algebraic varieties $V$ and $W$ embedded in a common ambient space is the variety
\begin{equation*}
V*W := \overline{\{ \lambda v + (1-\lambda)w\;|  \; \lambda \in \mathbb{C}, \; v \in V,\; w \in W\;\}}.
\end{equation*}
 In the special case when $V = W$, the join
 variety $V * V$ is called the \textit{secant variety of $V$}. Similarly, given two ideals $I_1, I_2 \subset \mathbb{C}[x_1, \dots, x_n]$, the ideal $I_1 * I_2  \subset \mathbb{C}[x_1, \dots, x_n]$ is the \textit{join ideal}. As for varieties, if $I_1 = I_2$, the ideal $I_1 * I_2$ is the \textit{secant ideal}. 
 We refer to \cite[Section 2]{Sturmfels2006} for the definition of $I_1 * I_2$, but note the following important property: 
\begin{equation}
\mathcal{I}_{\mathcal{T}_1} *\mathcal{I}_{\mathcal{T}_2} = I(V_{\mathcal{T}_1}*V_{\mathcal{T}_2}).  
\end{equation}

This package provides a means of computing invariants for those working in phylogenetic algebraic geometry.
We handle a class of commonly used models called 
\textit{group-based models} that have special restrictions 
on the entries of the transition matrices. 
They are subject to the Fourier-Hadamard coordinate transformation, which makes the parametrization monomial and the ideals toric
\cite{Evans1993, Szekely1993}. We will refer to the original coordinates, which represent leaf probabilities, as \emph{probability coordinates} and the transformed coordinates as \emph{Fourier coordinates}; furthermore, following the literature, we will use $p$ for probability coordinates and $q$ for Fourier coordinates. 
For these group-based models, we implement a theoretical construction for inductively determining the ideal of phylogenetic invariants for any tree from the invariants for claw trees \cite{Sturmfels2005}. 
We also handle the joins and secants of these ideals, 
which allows for computations involving mixture models.\\

\begin{figure}
\centering
\includegraphics[width=1.25in]{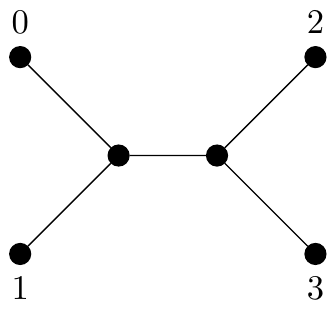}
\caption{Four leaf tree}\label{fig:4leaftree}
\end{figure}

\noindent
{\bf Functionality for Toric Phylogenetic Varieties.} As an example, let $\mathcal T$ be the four-leaf tree illustrated in Figure \ref{fig:4leaftree} and consider the Cavender-Farris-Neyman (CFN) model, a two-state group-based model, on $\mathcal T$. 
Then the toric ideal $\mathcal{I} _{\mathcal T}$ is generated  in degree $2$.  
Using {\tt PhylogeneticTrees.m2}, we can compute a generating set for $\mathcal{I} _{\mathcal T}$ using two different methods,
{\tt phyloToric42} and {\tt phyloToricFP}.

The first method {\tt phyloToric42} calls 
{\tt FourTiTwo.m2} \cite{M2}, 
the {\sc Macaulay2} interface to {\sc 4ti2} \cite{4ti2}, a software package with functionality for computing generating sets of toric ideals. 
We input $\mathcal{T}$ by its set of nontrivial splits.
In this instance, $01|23$ is the only
non-trivial split of $\mathcal T$, which we can enter as either
$\{0,1\}$ or $\{2,3\}$. The indices on the $q$'s correspond to two-state labelings of the four leaves of $\mathcal T$.\\

{\tt 
\begin{tabbing} 
Macaulay2, version 1.7  \\
\\
i1: \=load "PhylogeneticTrees.m2" \\
\\
i2: n = 4; T = \{\{0,1\}\}; M = CFNmodel; 
\\
\\
i5: toString phyloToric42(n,T,M) 
\\
\\
o5 =\>ideal(-q\_(0,1,1,0)*q\_(1,0,0,1)+q\_(0,
1,0,1)*q\_(1,0,1,0 
\\
\>),-q\_(0,0,1,1)*q
\_(1,1,0,0)+q\_(0,0,0,0)*q\_(1,1,1,1)) \\
\end{tabbing}}

The second method {\tt phyloToricFP} computes generators of $\mathcal I_{\mathcal T}$ using Theorem 24 in \cite{Sturmfels2005}; the ``FP" in the method name stands for fiber product 
\cite{Sullivant2007}.  For this example, this theorem allows us to explicitly construct a generating set of $\mathcal I_{\mathcal T}$ from generators of $\mathcal{I}_{K_{1,3}}$, the ideal associated to the CFN model on the claw tree $K_{1,3}$.  While our example here is binary, we note that this method is implemented for all trees, binary or not.\\
{\tt 
\begin{tabbing}
i6: \=toString phyloToricFP(n,T,M) \\
\\
o6 = \>ideal(
-q\_(0,0,1,1)*q\_(1,1,0,0)+q\_(0,0,0,0)*q\_(1,1,1,1), \\
\> q\_(0,0,1,1)*q\_(1,1,0,0)-q\_(0,0,0,0)*q\_(1,1,1,1),q\_(0,0, \\
\> 1,1)*q\_(1,1,0,0)-q\_(0,0,0,0)*q\_(1,1,1,1),-q\_(0,0,1,1)*q \\
\>\_(1,1,0,0)+q\_(0,0,0,0)*q\_(1,1,1,1),-q\_(0,1,1,0)*q\_(1,0, \\
\>0,1)+q\_(0,1,0,1)*q\_(1,0,1,0),q\_(0,1,1,0)*q\_(1,0,0,1)-q \\
\>\_(0,1,0,1)*q\_(1,0,1,0),q\_(0,1,1,0)*q\_(1,0,0,1)-q\_(0,1,0, \\
\>1)*q\_(1,0,1,0), -q\_(0,1,1,0)*q\_(1,0,0,1)+q\_(0,1,0,1)*q \\
\>\_(1,0,1,0))
\end{tabbing}}
\medskip

The algorithm used
by {\tt phyloToricFP} returns
more polynomials than are required to 
generate the ideal. If we wish to directly 
compare this ideal to that returned by 
{\tt phyloToric42} 
we must reconstruct both ideals 
in the same ring. Thus, we 
use the function {\tt qRing} to define
the ring of Fourier coordinates
and use the option of specifying the ring
for our ideals.
\\
{\tt 
\begin{tabbing}
i7: R = qRing(n,M) \\
\\
i8: \=phyloToric42(n,T,M,QRing=>R) == phyloToricFP(n,T,M,QRing=>R) \\
\\
o8 = \>true
\end{tabbing}}
\medskip

In our experiments, for most cases,
{ \tt phyloToric42} runs much faster
than { \tt phyloToricFP}.
This is likely because we have implemented
a naive version of 
the toric fiber product algorithm from 
\cite{Sturmfels2005}
with no attempt to 
avoid producing redundant
polynomials. It would be 
worth investigating if there is 
a faster implementation of this
algorithm. 
Still, one advantage offered by the fiber product method is the ability to inductively construct a single invariant when computing the entire ideal is infeasible. The method {\tt{phyloToricRandom}} returns such a randomly constructed invariant.

The polynomials that are returned by both methods are in Fourier coordinates, however, they can be converted to probability coordinates using the function {\tt fourierToProbability}. To do so, we must first construct the ring of probability coordinates using {\tt pRing}. 
Then the method {\tt fourierToProbability} returns a ring map that converts
polynomials in Fourier coordinates to probability coordinates. \\

{\tt 
\begin{tabbing}
i9: \= S = pRing(n,M); \\
\\
i10: \= phi = fourierToProbability(S,R,4,M);  \\
\\
i11: \= f = (vars R)\_(0,0) \\
\\
o11: \= q\_(0,0,0,0) \\
\\
i12: \= phi(f)\\
\\
o12: \=  (1/2)*p\_(0,0,0,0)+(1/2)*p\_(0,0,0,1)+(1/2)*p\_(0,0,1,0 \\
      \>)+(1/2)*p\_(0,0,1,1)+(1/2)*p\_(0,1,0,0)+(1/2)*p\_(0,1,0,1 \\ 
      \>)+(1/2)*p\_(0,1,1,0)+(1/2)*p\_(0,1,1,1)+(1/2)*p\_(1,0,0,0 \\
      \>)+(1/2)*p\_(1,0,0,1)+(1/2)*p\_(1,0,1,0)+(1/2)*p\_(1,0,1,1 \\
      \>)+(1/2)*p\_(1,1,0,0)+(1/2)*p\_(1,1,0,1)+(1/2)*p\_(1,1,1,0 \\
      \>)+(1/2)*p\_(1,1,1,1)\\
\end{tabbing}}


\noindent
{\bf Functionality for Secant Varieties}
Mixtures of group-based phylogenetic models correspond to secants and joins of toric ideals, objects that are of interest in combinatorial commutative algebra, but are notoriously hard to compute. 
In the methods {\tt joinIdeal} and {\tt secant}, 
we implement the elimination method described in 
\cite[Section 2]{Sturmfels2006} for computing the 
join of two homogeneous ideals or the 
secant of one homogeneous ideal.

 Consider now the Jukes-Cantor model on $\mathcal T$ from Figure \ref{fig:4leaftree}. The phylogenetic ideal for the mixture of 
 $\mathcal M_{\mathcal T}$ with itself is the 2nd secant ideal of
 the homogeneous ideal
 $\mathcal{I}_{\mathcal T}$, denoted 
 $\mathcal{I}_{\mathcal T} * \mathcal{I}_{\mathcal T}$.
 For secants, the method {\tt secant} takes as input a
 homogenous ideal and an integer $k$ 
 and returns a generating set for the $k$th secant ideal. 
 The method also accepts the optional
 argument {\tt DegreeLimit=>\{l\}}, 
 which computes generators of the ideal only up to degree $l$. 
 Thus, we can obtain generators of degree $3$ or less of $\mathcal{I}_{\mathcal T} * \mathcal{I}_{\mathcal T}$ 
 with the following commands.
The minimal generating set of {\tt SecI3} contains 49 linear 
invariants; we only print the generators with degree greater than one.\\

 {\tt
\begin{tabbing}
i13: \= I = phyloToric42(n,T,JCmodel); \\
\\
i14: \>SecI3 = secant(I,2,DegreeLimit=>{3}); \\
\\
i15: \>toString for i in flatten entries mingens SecI3 \\
\>list (if (degree i)\#0 == 1 then continue; i) \\
\\
o15 = \>
\{q\_(0,3,3,0)*q\_(3,0,2,1)*q\_(3,2,0,1)-q\_(0,3,2,1)*q\_(3, \\
\>0,3,0)*q\_(3,2,0,1)+q\_(0,3,2,1)*q\_(3,0,0,3)*q\_(3,2,1,0 \\
\>)-q\_(0,3,0,3)*q\_(3,0,2,1)*q\_(3,2,1,0)-q\_(0,3,3,0)*q\_(3 \\
\>,0,0,3)*q\_(3,2,3,2)+q\_(0,3,0,3)*q\_(3,0,3,0)*q\_(3,2,3,2 \\
\>)\}
\end{tabbing}}

The degree bound allows for the possibility of obtaining some invariants
when computing a generating set for the secant ideal is infeasible. 
In some instances, 
$(\mathcal{I}_{\mathcal{T}}*
\mathcal{I}_{\mathcal{T}})_{l}$, 
the ideal of generators of degree less than or 
equal to $l$, may in fact generate the entire ideal.
To prove this we 
must verify that
$\dim((\mathcal{I}_{\mathcal{T}}*
\mathcal{I}_{\mathcal{T}})_{l}) = 
\dim(\mathcal{I}_{\mathcal{T}}*
\mathcal{I}_{\mathcal{T}})$ 
and that
$(\mathcal{I}_{\mathcal{T}}*
\mathcal{I}_{\mathcal{T}})_{l}$ 
is prime.
Assuming we are able to compute 
$(\mathcal{I}_{\mathcal{T}}*
\mathcal{I}_{\mathcal{T}})_{l}$, we can
compute its dimension and verify that it is 
prime.
We then know that 
$\dim((
\mathcal{I}_{\mathcal{T}}*
\mathcal{I}_{\mathcal{T}})_{l}) 
\geq
\dim(
\mathcal{I}_{\mathcal{T}}*
\mathcal{I}_{\mathcal{T}})$,
leaving the inequality
$\dim((
\mathcal{I}_{\mathcal{T}}*
\mathcal{I}_{\mathcal{T}})_{l}) 
\leq
\dim(
\mathcal{I}_{\mathcal{T}}*
\mathcal{I}_{\mathcal{T}})$ to show.
The method {\tt toricSecantDim} enables us 
to do this using
a probabilistic method based on Terracini's Lemma \cite{Terr1911}
to compute a lower bound on 
$\dim(
\mathcal{I}_{\mathcal{T}}*
\mathcal{I}_{\mathcal{T}}
)$. 

Using this method, we can show that the secant
of the ideal from the previous example is in fact
generated in degree less than three. 
\\

 {\tt
\begin{tabbing}
i16: \= dim(SecI3) \\
\\
o16 \>= 12 \\
\\
i17: \= isPrime(SecI3) \\
\\
o17 \>= true \\
\\
i18: \= toricSecantDim(phyloToricAMatrix(4,\{\{0,1\}\},JCmodel),2)) \\
\\
o18 \>= 12
\end{tabbing}}

\medskip 

In the code above, we used 
{\tt phyloToricAMatrix(n,T,JCmodel)} 
to construct the $\mathcal{A}$ matrix of 
the toric ideal. For more details 
see the documentation and \cite{sturmfels1996grobner}.
In this instance, the method outlined is
substantially faster than using the
{\tt secant} method without a degree bound. \\
\\



\noindent {\bf Additional functionality:} Although this package was developed with toric ideals from phylogenetics in mind,
the methods {\tt secant} and {\tt joinIdeal}
can be used for {\it any} homogeneous ideals. 
Thus, these can be employed for computations 
outside of phylogenetic algebraic geometry.

The following models are loaded with the package: the Cavender-Farris-Neyman model (CFNmodel), the Jukes-Cantor model (JCmodel), the Kimura 2-parameter model (K2Pmodel), and the Kimura 3-parameter model (K3Pmodel). Additionally, some functionality for working with trees is available in this package, which includes the methods 
{\tt internalEdges},
{\tt internalVertices},
{\tt edgeCut},
{\tt vertexCut},
{\tt edgeContract}.\\
\\
{\bf Acknowledgements.} This work began at the 2016 AMS Mathematics Research Community on ``Algebraic Statistics," which was supported by the National Science Foundation under grant number DMS-1321794. RD was supported by NSF DMS-1401591. EG was supported by NSF DMS-1620109.
RW was primarily supported by a NSF GRF under grant number PGF-031543 and partially supported by the NSF RTG grant 0943832.

\bibliography{references}
\bibliographystyle{plain}

$\ $\\

\end{document}